\documentclass[crop]{CSL}%%%%where CSL is the template name

\usepackage{amsmath} %for dealing with mathematics,

\bibpunct[]{(}{)}{;}{a}{}{,}

\begin{document}

\supertitle{Research Paper}

\title[recto running head]{Solid grains ejected from terrestrial exoplanets as a probe of the abundance of life in the Milky Way}

\author[verso running head]{Tomonori TOTANI$^{1,2}$}

\address{\add{1}{Department of Astronomy, School of Science,  
The University of Tokyo, 
7-3-1 Hongo, Bunkyo-ku, Tokyo 113-0033, Japan} 
and \add{2}{Research Center for the Early Universe, School of Science, The University of Tokyo, 
7-3-1 Hongo, Bunkyo-ku, Tokyo 113-0033, Japan }}

\corres{\name{Tomonori TOTANI} \email{totani@astron.s.u-tokyo.ac.jp}}

\begin{abstract}
Searching for extrasolar biosignatures is important to understand life on Earth and its origin.
Astronomical observations of exoplanets may find such signatures,
but it is difficult and may be impossible to  
claim unambiguous detection of life by remote sensing of exoplanet atmospheres. 
Here, another approach is considered: collecting grains ejected by asteroid impacts
from exoplanets in the Milky Way and then traveling to the Solar System. 
The optimal grain size for this purpose is around 1 $\mu$m, and though uncertainty is large, 
about $10^5$ such grains are expected to be accreting on Earth every year,
which may contain biosignatures of life that existed on their home planets.
These grains may be collected by detectors placed in space, or
extracted from Antarctic ice or deep-sea sediments, depending
on future technological developments. 
\end{abstract}


\selfcitation{}

\received{xx xxxx xxxx}

\revised{xx xxxx xxxx}

\accepted{xx xxxx xxxx}

\maketitle

\Fpagebreak

%--- main text starts

\section{Introduction}

The abundance of life in the universe is poorly known \citep{Lineweaver2002,
Spiegel2011, Kipping2020}. In the Drake equation \citep{Vakoch2015},
the probability $f_l$ of life in any form appearing on 
a habitable planet may not be of order unity. Instead, it may be so small that 
Earth is the only planet that harbors life
in the observable universe (13.8 billion light-year radius in light travel distance), 
though many abiogenesis events may have occurred in 
the total volume of an inflationary universe \citep{Totani2020}. 
Knowing $f_l$ quantitatively is crucially important to get a hint about the completely
unknown processes of abiogenesis, and it can be
constrained by searching for extraterrestrial signatures of life. 
If there is no life of a different origin than ours
in the Solar System, we need to search for extrasolar biosignatures, which
can be probed by astronomical observations of
nearby exoplanet atmospheres \citep[see][ for a review]{Kiang2018}.
However, any candidate biosignature is likely to be controversial,
given that even oxygen (a representative biomarker in exoplanet atmospheres)
can be generated abiologically 
\citep{Meadows2018,Bains2021, Meadows2022,Smith2022}.
SETI (search for extra terrestrial intelligence) may find unambiguous extrasolar
technosignatures, but it is sensitive only to intelligent life, which may be
much less abundant than primitive life forms. 
Thus it is uncertain whether we will obtain an irrefutable constraint on $f_l$
in the future by methods proposed so far. Here, another approach of
directly sampling extrasolar biosignatures of primitive life or microbes
is considered. 

\section{Ejection of rocks and grains from terrestrial planets}

Meteorites of Martian origin found on Earth demonstrate that material has been
exchanged between planets in the Solar System. These rocks are thought
to have been ejected from their home planets by giant impact events of asteroids. 
The possibility of living microbes on such ejecta migrating to other planets has been 
discussed in the literature as the panspermia hypothesis
\citep[][ for a review]{Nicholson2009}.
Rocks ejected from Earth or Mars eventually fall onto planets after 
$\sim 10^{6-7}$ yrs stay in orbits, but 10--20\% of these are ejected
from the Solar System by interaction with the giant planets \citep{Gladman2000, Melosh2003}.
This implies the possibility of interstellar transfer of living organisms, 
but previous studies found negligibly small success probabilities
\citep{Melosh2003, Wallis2004, Valtonen2008, Adams2022}. 
Interstellar panspermia is limited not only by large distances to nearby stars, 
but also by the survival time of organisms in space 
($\sim 10^{5-7}$ yrs, \citealt{Valtonen2008, Wesson2010}) 
and the minimum rock size ($\gtrsim 10$ kg, \citealt{Valtonen2008, Adams2022}) to protect
microbes during the journey.
 
However, if our purpose is to capture 
these ejecta particles after traveling to Earth as a search
of extrasolar biosignatures, these constraints do not apply\footnote{
\citet{Cataldi2017} considered the possibility of remotely detecting these ejecta
bound in their home planetary system by astronomical observations from Earth.}. We do not
need living organisms, but the existence of life in exoplanets can be probed by
remains of microbes, microfossils, 
minerals produced by biological activities (biominerals), or any other
signatures of past biological activities (e.g. concentration of biological molecules or
isotopic ratios) \citep{Cavalazzi2018}. The mineralogical diversity of Earth is larger
than Venus or Mars, which is thought to be due to the presence of life
\citep{Hazen2008,Cataldi2017}. Then smaller rocks or grains
are more favorable, because particle flux would be larger and hence
a higher detection probability is expected. 
It is reasonable to consider grains larger
than the minimum size of bacteria ($\sim 1 \mu$m), so that
direct biosignatures (remains or microfossils) can be contained. 

Typically, the total mass of ejecta in a single meteorite impact is about 
$10^{-3}$ of the impactor mass, and total mass ejection rate from a 
terrestrial planet has been estimated at $10^{4-5}$ kg/yr
in the Solar System for ejecta mass larger than $\sim 10$ kg
\citep{Wallis2004, Napier2004, Valtonen2008}. 
It should be noted that these estimates were made for panspermia studies 
and take into account the conditions under which microorganisms can survive the launch
(temperatures $< $ 100 C$^\circ$ and shock pressure $P < 1$ GPa).  
The estimate is likely to increase for our purpose because the weaker condition 
(preservation of biosignatures) is sufficient. If we impose only the condition of
nonmolten ejecta, the estimate may increase by a factor of 10--60 \citep{Cataldi2017}.
Here we conservatively present the following calculations based on
the $10^{4-5}$ kg/yr rate. Small ejecta particles may not escape the planet
if they are decelerated as they pass through the atmosphere. However,
the impact would produce high-velocity vapor and particle plume which blows a hole
through the atmosphere, and then small particles may also be ejected 
\citep{Melosh1988, Wallis2004}. 

Mass/size distribution of ejected rocks or grains, $dN/dm$
(number of ejecta per unit ejecta mass $m$), is expected to be broad. 
When rocks on a planet's surface break up by dynamic fragmentation following
an impact, fragment mass distribution is $dN/dm \propto m^{-\alpha}$
with $\alpha \sim 2$ \citep{Melosh1992}. When a rock is ejected into space,
smaller grains or soils can also be pushed out by the rock, whose total mass is comparable
to the rock. The mass/size distribution may further be altered by 
collisions in interplanetary space. Since
the collisional disruption time is shorter than $10^{6-7}$ yrs for grains smaller than 1 cm 
\citep{Leinert1983},  they are converted into smaller grains during the
typical residence time in interplanetary space.  The mass distribution of asteroids
is a result of such collisional cascades, and it is also a power-law with
$\alpha \sim 2$ \citep{Dermott2001}.  Therefore, it is reasonable
to expect an ejecta supply rate $d\dot N/dm$ into interplanetary space 
to be a power-law with $\alpha \sim 2$.  In this case, the total ejecta mass
per logarithmic interval of $m$, $m \, d\dot N/d(\ln m) = m^2 \, d\dot N/dm$
is constant against $m$. Then, in a wide range of $m$, the total ejecta
mass supply rate is $\sim10^4$ kg/yr in an interval of $m_l < m < m_u$ with
$m_u/m_l = e = 2.718...$, and it would be increased by a factor of at most 10 when 
a wider mass range (e.g., $m_u/m_l = 10^5$) is considered.

\section{Travels in interplanetary and interstellar space}

It is necessary to examine whether micron-sized grains 
can be ejected from the Solar System, because such small grains are
affected by non-gravitational processes in interplanetary space, 
in contrast to larger bodies \citep{Dermott2001, Koschny2019}.  
Magnetic fields are not important for grains larger than 1 $\mu$m, 
but the Poynting-Robertson (PR) drag time is
shorter than the collisional time for grains smaller than 100 $\mu$m.
Then particles smaller than 1 cm will spiral into the Sun within $10^6$ yrs by collisional 
disruption and subsequent PR drag, but 
collisions are still effective during the inspiral to produce even smaller micron-sized grains. 
Solar radiative pressure is greater compared to gravity for smaller grains,
and the two forces are comparable at $\sim 1 \ \mu$m
size. Grains of this size are then ejected from the Solar System as ``beta meteoroids''
\citep{Zook1975, Dermott2001}, and consequently
a significant fraction of originally 1--$10^4 \ \mu$m ejecta will eventually escape
from the Solar System as micron-sized grains \citep{Leinert1983, Napier2004, Wesson2010}. 
In the following, it is assumed that 
the Solar System is typical among planetary systems in the Milky Way,
and grains about  $1 \ \mu$m in diameter ($m \sim 10^{-12}$ g
for a spherical grain of density $\sim 3 \ \rm g \ cm^{-3}$)
that were originally on a planet's surface are ejected from a planetary system
into interstellar medium (ISM) at a total mass ejection
rate of $\dot M_{\rm ej} = 10^4$ kg/yr. 

Micron-sized grains may be damaged or destroyed during interstellar travel.
Conventional theoretical estimates of the lifetime of
interstellar dust particles are relatively short ($\sim$ 0.1--1 Gyr, \citealt{Jones1997}).
However, these estimates are highly uncertain, and such short lifetime
conflicts with the observed dust abundance and dust production time
scales in the Milky Way as well as distant galaxies
\citep{Jones2011, Rowlands2014, Ferrara2021}. About ten times longer lifetime 
is then favorable, and some theoretical studies support this possibility
\citep{Slavin2015, MartinezGonzalez2019}. Furthermore,
micron-sized grains considered here are larger than the general
interstellar dust grains responsible for the extinction of light from astronomical objects  
(5--250 nm, \citealt{Mathis1977, Grun2000}).
Such large grains are decoupled from ISM gas and smaller grains, resulting in
a longer lifetime \citep{Grun2000, Frisch2003, Slavin2004, Hirashita2016}. 
Therefore, the interstellar lifetime of micron-sized grains, $\tau_{is}$, 
could be comparable with the age of the Solar System or
the Milky Way ($\tau_{is} \sim$ 1--10 Gyr), though further studies are
necessary for a more quantitative estimate.

\section{Exoplanetary grain flux to Earth}

Now we can estimate the interstellar density 
$\rho_{ej} = \dot M_{ej}  \, n_* \, f_{hp} \, \tau_{is} $ of the micron-sized
grains ejected from habitable terrestrial planets in the Milky Way, 
where $n_*$ is the number density of Sun-like stars and $f_{hp}$ is the
fraction of stars having a habitable planet. Here, the spatial distribution
of the grains is assumed to be the same as that of stars, and the evolution of the Milky Way
(e.g., star formation history) is not taken into account, but such a 
simple treatment is sufficient for this work. 
Adopting  $n_* = 0.03 \ \rm pc^{-3}$ (for stars in the solar neighborhood 
heavier than $\sim 0.3 M_\odot$, \citealt{Bovy2017}), 
$f_{hp} = 0.1$ \citep{Lissauer2014}, and $\tau_{is}$ = 10 Gyr,
$\rho_{ej}$ is found to be $1.0 \times 10^{-41} \ \rm g \ cm^{-3}$,
which is $\sim 10^{14-15}$ times smaller than dust density
in typical ISM of
a hydrogen number density $n_H \sim 0.3 \ \rm cm^{-3}$ \citep{Grun2000}
and a dust-to-gas mass ratio of 0.01 \citep{Tricco2017}. 
The radial migration of stars in the Milky Way disk is estimated to be more than 1 kpc
for a time scale of a few Gyrs \citep{Frankel2018, Lian2022}, and hence the grains observed
at one location is a mixture of a considerable fraction of the Galaxy,
in contrast to astronomical observations of nearby exoplanets. 
 
The density $\rho_{ej}$
can be converted into particle flux $F_{ej} = \rho_{ej} \, \langle v \rangle / (4 \pi m)$
per steradian, where $\langle v \rangle$ is the mean velocity of grains. Motion of micron-sized
grains are affected by solar radiation when entering the Solar System, but
{\it in situ} spacecraft measurements indicate that the flux of interstellar
dust grains of $m \sim 10^{-12} \ \rm g$ 
is not reduced at a distance of about 1 au from the Sun \citep{Grun1997}. 
Then the micron-sized
grains from terrestrial exoplanets are accreting on Earth directly from ISM at a rate of 
$A_\oplus = 4 \pi F_{ej} \, \pi \, r_\oplus^2 \sim 1.6 \times 10^3$ particles every year,
where $r_\oplus$ is the Earth radius. Here,  $m = 10^{-12}$ g and
$\langle v \rangle = 40$ km/s (for the Maxwell distribution
with one-dimensional velocity dispersion $\sigma_v = 25$ km/s, \citealt{Cox2013})
are adopted, assuming that the grain velocity distribution in ISM is the same as
stars in the solar neighborhood.

The rate $A_\oplus$ may further be increased if we consider gravitational capture
of grains by interaction with giant planets in the Solar System. 
Mainly low-velocity objects are captured and bound to the Solar System
by a cross section
$\sigma_c (v) = \sigma_0 \, (v/v_c)^{-2} \, [ 1 + (v/v_c)^2]^{-2}$,
where $\sigma_0 = 2.32 \times 10^5 \ \rm au^2$ and $v_c = 0.42$ km/s \citep{Adams2022}. 
A fraction $f_\oplus \sim 10^{-4}$ of the
captured objects are expected to hit Earth before they are lost from interplanetary space
\citep{Melosh2003, Adams2022}. 
Then the accretion rate $A_\oplus$ is enhanced by a factor of 
$\eta_c = \langle \sigma_c v \rangle f_\oplus / (\pi r_\oplus^2 \langle v \rangle)
\sim 10^2$ compared with direct hitting from ISM, and
thus $A_\oplus \sim 1.6 \times 10^5 \,
\eta_{c,2}$ particles per year, where $\eta_{c,2} \equiv \eta_c / 10^2$. It should be noted
that the estimate of $\eta_c$ (especially $f_\oplus$) is highly uncertain, 
which is considering only
gravitational interactions. Radiative pressure and the PR drag effect may change
this estimate. 

These grains from terrestrial exoplanets can be collected by detectors 
placed in space, utilizing a low-density capture media like silica aerogel that
enables capture of hypervelocity particles with mild deceleration 
and hence minimal damage to biosignatures \citep{Westphal2014, Yamagishi2021}. 
A very large total effective area
(hopefully comparable to Earth, or $\sim 10^3 \ \rm km^2$
to expect one particle detection per year) is
necessary to detect these particles, but it may be possible in the future, 
depending on technological developments and humanity's advance into space. 
A single large detector is not necessary, but a large number of small and low-cost
detectors would be more realistic. Large space telescopes in the next generation
cost more than 10 billion USD \citep{ASTRO2020}, 
targeting indirect biosignatures
in exoplanet atmospheres. It would be interesting to think about what can be done
at the same cost to search for direct biosignatures by collecting grains from exoplanets. 

Another collecting method may be to search on Earth for these exoplanet grains. 
Cosmic dust 
particles smaller than 10--100 $\mu$m survive atmospheric entry without
severe heating \citep{Koschny2019}, and hence biosignatures are not seriously damaged
by the entry process. 
Such micrometeorites have been collected from Antarctic snow or ice
\citep{Yada2004, Rojas2021}. The accretion rate $A_\oplus$ estimated above
implies that $10^9 \, \eta_{c,2}$ grains from exoplanets are
embedded in the entire Antarctic ice ($1.4 \times 10^7 \ \rm km^2$ area
and 2500 m mean depth corresponding to an accumulation time of
$\sim 3 \times 10^5$ yr, \citealt{Kawamura2017}), or $10^2 \, \eta_{c,2}$ 
grains in an area of $1 \ \rm km^2$. Cosmic dust particles have also been
collected in deep-sea sediments, where low accumulation rates and
long exposure times allow extraterrestrial particles to collect in
high concentrations \citep{Brownlee1985}. For a typical accumulation
rate ($2 \times 10^{-6}$ m/yr), about $10^4 \, \eta_{c,2}$ grains from
exoplanets can be collected from deep-sea clay of 1 m depth
(corresponding to $5 \times 10^5$ yr) in a $100 \ \rm km^2$
area. It is worth investigating the best locations and strategies to collect these extremely rare
particles on Earth by future technologies.

\section{Uncertainties and issues for future consideration}

It is clear that there are large uncertainties (probably a few orders of magnitude or more
in total) in the flux estimate of exoplanet dust particles 
in this work, at various stages of launch from the home planets,
escape from the home exoplanetary systems, travel in interstellar space, and
capture by the Solar System and by Earth. However, the estimate is large enough
to make the future detection of such exoplanetary particles a realistic possibility, and 
merits further study. 

Another important issue not discussed here
is whether biosignatures are preserved until the exoplanet particles reach Earth.
They may be damaged at various stages, including launches from the home planets,
exposure to radiation and cosmic rays in interstellar space, entry to Earth, and
weathering in Earth environments. Microbial carcasses would be most vulnerable to damage,
while microfossils and biominerals would be more likely to be preserved.
It is important to investigate and choose 
the best biosignatures for this purpose, which should be abundant 
on terrestrial planets harboring life
and identifiable after a long travel from their home. 

Identifying grains of extrasolar origin would not be easy after eliminating 
the possibility of terrestrial or Solar-System origin. 
Grains detected directly from interstellar space
may be identified by their orbits. Extrasolar particles scattered by giant planets and then
bound to the Solar System may be difficult to distinguish from particles ejected from Earth,
even if they contain biosignatures.
Long residence times in interstellar space inferred from cosmic ray and radiation exposure
would be useful, because grains ejected from Earth would be lost in $\sim 10^7$ yrs
like interplanetary dust particles. 
Identification would be even more difficult for particles collected on Earth. 
Extraordinary biosignatures that are quite different from known Earth life, as well as
anomalous isotope ratios and/or mineralogical compositions,
are expected to be helpful in identifying biosignatures of extrasolar particles.
Finding even just one such particle would have an immense
impact on the origin of life studies. 

More quantitative considerations are beyond the scope of this paper and require experimental 
studies by experts in various fields. Given the possibility of getting biosignatures in
direct samples from exoplanets, further research in this direction is recommended.

%--- main text ends

%\ack[Acknowledgement]{Insert the Acknowledgment text here.}

\ack[Conflict of interest]{None.} 

\bibliographystyle{apa}
\bibliography{totani2022_IJA}

%%%%%%%%%%%% Biography text %%%%%%%%%%%%%%%%%%%%%%%%%%

\end{document}